\documentclass[showpacs,twocolumn,prl]{revtex4}
\usepackage{graphicx}
\begin{document}
\title{Spin-charge-lattice coupling near the metal-insulator transition in Ca$_3$Ru$_2$O$_7$\\}

\author{C.S. Nelson}
\affiliation{National Synchrotron Light Source, Brookhaven National Laboratory, Upton, NY  11973-5000, USA}
\author{H. Mo}
\affiliation{National Synchrotron Light Source, Brookhaven National Laboratory, Upton, NY  11973-5000, USA}
\author{B. Bohnenbuck}
\affiliation{Max-Planck-Institut f\"{u}r Festk\"{o}rperforschung, Heisenbergstra\ss e 1, D-70569 Stuttgart, Germany}
\author{J. Strempfer}
\affiliation{Hamburger Synchrotronstrahlungslabor HASYLAB at Deutsches Elektronen-Synchrotron DESY, Notkestra\ss e 85, 22603 Hamburg, Germany}
\author{N. Kikugawa}
\affiliation{National Institute for Materials Science, Tsukuba, Ibaraki 305-0047, Japan}
\author{S.I. Ikeda}
\affiliation{National Institute of Advanced Industrial Science and Technology, Tsukuba, Ibaraki 305-8568, Japan}
\author{Y. Yoshida}
\affiliation{National Institute of Advanced Industrial Science and Technology, Tsukuba, Ibaraki 305-8568, Japan}

\date{\today}
\begin{abstract}
We report x-ray scattering studies of the c-axis lattice parameter in Ca$_3$Ru$_2$O$_7$ as a function of temperature and magnetic field.  These structural studies complement published transport and magnetization data, and therefore elucidate the spin-charge-lattice coupling near the metal-insulator transition.  Strong anisotropy of the structural change for field applied along orthogonal in-plane directions is observed.  Competition between a spin-polarized phase that does not couple to the lattice, and an antiferromagnetic metallic phase, which does, gives rise to rich behavior for {\bf B} $\parallel b$.
\end{abstract}
\maketitle

The interplay of spin, charge, and lattice degrees of freedom is believed to underlie the complicated physics exhibited by many correlated electron systems, such as the colossal magnetoresistant manganites.\cite{kaplan}  An exciting corollary of this interplay is the tunability of ground state properties--- e.g., inducing a metal-insulator transition via application of magnetic field, or controlling magnetic properties via strain--- that has the potential to turn these systems into technologically useful electronic materials.\cite{bishop}  Optimization of the useful properties, however, requires a thorough understanding of the spin-charge-lattice coupling in these materials.

Bilayer Ca$_3$Ru$_2$O$_7$ provides a particularly rich example of spin-charge-lattice coupling in a correlated electron system.  Metallic Ca$_3$Ru$_2$O$_7$ orders antiferromagnetically at T$_N$ = 56 K and exhibits a metal-insulator transition at T$_{m-i}$ = 48 K,\cite{cao1} although a quasi-2D metallic ground state has been reported for T $<$ 30 K.\cite{yoshida1}  At T$_{m-i}$, a collapse of the c-axis lattice parameter is observed,\cite{cao3} with a concomitant in-plane expansion.\cite{yoshida2}  This structural change is surprising in that a smaller c-axis lattice parameter is expected to increase the orbital overlap, which would result in a stabilization of the metallic state.  Neutron powder diffraction studies suggest, however, that a decrease in the Ru-O-Ca bond angle that decreases the interlayer electron transfer may explain this unexpected behavior.\cite{yoshida2}  Finally, a spin reorientation transition, which was indicated by magnetic susceptibility measurements,\cite{cao1} occurs in the vicinity of T$_{m-i}$.\cite{mccall,cao4,yoshida1,bohnenbuck}  Neutron powder diffraction\cite{yoshida2} confirmed the proposed low-temperature magnetic structure\cite{mccall} as consisting of ferromagnetic bilayers, antiferromagnetically coupled, and with moments along the orthorhombic {\it b} axis, using the {\it Bb}2$_1${\it m} notation of space group \#36, for which {\it a} $<$ {\it b}.\cite{abswitch}  The spin reorientation rotates the moments such that they lie along {\it a} for T $>$ T$_{m-i}$.  

The confluence of spin reorientation, structural change, and a metal-insulator transition clearly demonstrates the interplay of the spin, charge, and lattice degrees of freedom in Ca$_3$Ru$_2$O$_7$ near T$_{m-i}$, which has been the focus of recent transport and magnetization measurements.\cite{mccall,lin}  In this paper, we further this investigation by reporting x-ray scattering studies of the c-axis lattice parameter in Ca$_3$Ru$_2$O$_7$, with applied magnetic fields of up to 10 T.  Combining this new information with previously reported measurements results in a more complete picture of spin-charge-lattice coupling near the metal-insulator transition in this material.  We observe a sharp structural change for magnetic field applied along the low-temperature hard axis that is driven by strong magnetoelastic coupling, and ties the reported colossal magnetoresistance\cite{cao4} to an increase in the c-axis lattice parameter.  For magnetic field applied along the low-temperature easy axis, we observe a gradual change in the c-axis lattice parameter as a function of temperature, and irreversibility as a function of magnetic field.  Correlating this behavior with transport and magnetization data suggests competition between the spin-polarized and antiferromagnetic metallic phases.

Single crystal samples were grown at the University of St. Andrews and at the National Institute of Advanced Industrial Science and Technology (AIST), using floating zone techniques.  Detailed information about the growth techniques and transport behavior of the AIST-grown samples have been reported elsewhere.\cite{yoshida1}  The as-grown samples are shaped like platelets, with the c-axis along the short direction, and relatively flat (001) surfaces.  We note that samples grown at the two institutions are observed to be very similar:  the mosaic widths, as characterized at the (004) reflections, are 0.1--0.2$^{\circ}$, and the temperatures at which the zero field structural change are observed differed by $\sim$0.2 K.  Specifically, the structural changes, which are concomitant with the metal-insulator transition, are at 47.92 $\pm$ 0.10 K and 48.14 $\pm$ 0.10 K for the St. Andrews and AIST samples, respectively.  For ease of comparison in what follows, the latter value (T$_{m-i}$ = 48.14 K) is used, and 0.22 K has been added to the measured sample temperature for all data collected from the St. Andrews sample.  In addition, the {\it Bb}2$_1${\it m} notation of space group \#36 is used, and therefore the low-temperature easy(hard) axis is the {\it b}({\it a}) axis.\cite{abswitch} 

X-ray scattering measurements were carried out on wiggler beamline X21 at the National Synchrotron Light Source.  A Si(111) double-crystal monochromator was used to select the incident energy, and energies of 8.9 and 12 keV were used during different experimental runs.  The monochromatic beam was focused to a $\sim$1 mm$^2$ spot at the center of rotation of a 2-circle goniometer.  Mounted on top of this goniometer were {\bf x}, {\bf y}, and {\bf z} translation stages, $\pm 4^{\circ}$ orthogonal tilt stages, and a 13 T, split coil, vertical field, superconducting magnet, which was made by Oxford Instruments.  Scattering was carried out in a horizontal geometry, and an avalanche photodiode was used as the detector.  The platelet-shaped sample was glued to a brass holder attached to the end of a sample rod such that the c-axis was in the scattering plane (i.e., {\bf Q} $\parallel c^*$).  The holder enabled a manual rotation of the sample about the (001) surface normal, and therefore the magnetic field could be applied along either the {\it a} or {\it b} in-plane axis (see inset to Figure 1).

After initially cooling the sample to $\sim$25 K, zero field measurements were carried out while increasing the sample temperature.  In Figure 1, longitudinal $\theta$-$2\theta$ scans through the (0 0 16) reflection are shown as a function of temperature as $T_{m-i}$ is approached.  A clear shift to lower Q due to the c-axis lattice parameter expansion is observed between the temperatures of 47.8 and 48.55 K.  Although the magnitude of the shift is less than the full-width-at-half-maximum of the (0 0 16) reflection, Gaussian fits to these data result in a peak position value with $\sigma < 0.002^{\circ}$.  This translates into a sensitivity to the change in the c-axis lattice parameter of $\sim$0.002\% at this Q.

The field dependence of the structural change was investigated for magnetic field applied along the {\it a} and {\it b} in-plane directions.  For all fixed-B measurements, the sample was cooled in zero field to a temperature of $\sim$25 K, the field was ramped up to its final value, and data were collected at fixed B while increasing the sample temperature.  These fixed-B data sets are summarized in Figure 2, in which the percent change in the c-axis lattice parameter from its value at  T $\approx$ 25 K is plotted as a function of temperature.  For both field orientations, the temperature at which the structural change begins is observed to decrease with increasing field, which indicates that magnetic field stabilizes the high-temperature structure.  However, both the magnitude of the shift, and more dramatically, the temperature range over which the c-axis lattice parameter changes, exhibit strongly anisotropic behavior for the two field orientations.  Focusing first on Figure 2(a) in which {\bf B} $\parallel a$, a step-like change in {\it c} of roughly constant magnitude is observed for B $\leq$ 5 T.  For B = 8 T, the magnitude of the step is reduced, which is presumably related to the sample only being cooled to $\sim$25 K:  the c-axis resistivity at T = 25 K is reported to decrease from its zero field value for B $\geq$ 7 T,\cite{cao5} and the strong charge-lattice coupling that will be discussed below suggests that a change in the structure is concomitant with a change in the transport.  In Figure 2(b) for {\bf B} $\parallel b$, the step-like change in {\it c} is absent, and a transition temperature is difficult to determine.  For B = 8 T this is not unexpected since 8 T is greater than the critical field for the metamagnetic transition observed for {\bf B} $\parallel b$.\cite{cao2a}  The high-temperature asymptotic approach of the B = 8 T data set to the other data sets indicates that there is no significant change in the c-axis lattice parameter as the field is ramped up to 8 T, and therefore that the metamagnetic transition to a spin-polarized phase is not strongly coupled to the lattice.  Note that this conclusion is also supported by magnetostriction\cite{ohmichi} and Raman scattering\cite{karpus} studies.

The field dependence of the structural change shown in Figure 2 can be used to extract a B-T phase diagram.  For {\bf B} $\parallel a$, the temperature at which the step-like change is observed is plotted in closed symbols in Figure 3 to indicate the phase boundary.  As mentioned above, for {\bf B} $\parallel b$ a clear phase boundary cannot be determined given the nature of the change in {\it c}.  Therefore two temperatures--- indicative of the beginning and the end of the structural change for applied fields less than the critical field for the metamagnetic transition ($\sim$6 T)--- have been extracted from each of the data sets shown in Figure 2(b), and plotted as the open symbols in Figure 3.  These beginning and ending points mark the zero-field jump in the c-axis lattice parameter, and are indicated by the dashed lines in Figure 2(b).

It is illuminating to compare the structural change phase diagram of Figure 3 with the B-T phase diagrams for c-axis transport and magnetization reported by Cao and co-workers.\cite{mccall,lin}  After noting that the in-plane lattice parameters in these papers are reversed with respect to the designation used in this paper,\cite{abswitch} the structural change phase boundary is seen to be very similar to the phase boundaries determined through transport.  Specifically, the temperature at which colossal magnetoresistance is observed for {\bf B} $\parallel a$ coincides with the structural change phase boundary for T $< T_{m-i}$, which indicates an intimate coupling between the structure and transport.  For {\bf B} $\parallel b$, the structural change also appears to follow the phase boundary from transport, although only for the dominant resistivity change (i.e., $B_{c1}$, not $B_{c2}$, in Lin {\it et al.}\cite{lin}).  Note that this is true not merely for T $< T_{m-i}$, but that the field dependence of the temperature corresponding to the end of the structural change mimics the behavior of a second phase boundary determined through transport for T $> T_{m-i}$.  This second phase boundary is marked by a subtle inflection in the c-axis resistivity at which no change in the magnetization is observed, and was suggested to be a manifestation of the crucial role of the orbital degree of freedom.\cite{lin}  Our structural measurements indicate that, in fact, the lattice degree of freedom may be more relevant.  Taken together, our results demonstrate a strong charge-lattice coupling over the field and temperature range investigated, and independent of the orientation of the in-plane field.

One obvious question raised by the phase diagram of Figure 3 is:  what is the nature of the phase between the two dashed lines, for {\bf B} $\parallel b$?  From our fixed-B measurements we know that the c-axis lattice parameter gradually increases with increasing temperature in this phase.  Fixed-T measurements can add to our understanding by indicating how magnetic field alone affects the structure.  For these measurements, the sample was again cooled in zero field to a temperature of $\sim$25 K, the sample temperature was then increased to the temperature of interest, and data were collected at fixed T while ramping up the field.  In order to correct for a shift in the position of the rod with increasing field, which moves the sample away from the center of rotation and therefore shifts the $2\theta$ position of the Bragg peaks slightly, data collected at T = 70 K was used to measure the rod shift that was then subtracted from all T $< T_{m-i}$ measurements.  A result of one such fixed-T measurement--- with T = 44.3 K and {\bf B} $\parallel$ {\it b}--- is displayed in the inset to Figure 3.  The structural change is observed to occur between field values of 3--5 T, which is consistent with the leading edge phase boundary of the structural change in Figure 3.  Intriguingly, the total change saturates at a mere 0.04\%, which is a factor of $\sim$2 less than the change as a function of temperature in zero field.  Another interesting feature of these data can be noted from a comparison with the fixed-B data sets shown in Figure 2(b), in which $\Delta c$ at T = 44.3 K is much larger for B = 5 T than for B = 8 T.  In the inset to Figure 3 this is not the case, which suggests that the path taken to the T = 44.3 K and B = 8 T point in phase space matters.  This will be discussed in more detail below.

Additional fixed-T measurements were carried out for both field orientations, and hysteresis curves from these measurements are shown in Figure 4.  The {\bf B} $\parallel a$ data set measured just below $T_{m-i}$ displays sharp transitions with no obvious hysteresis, and at a field consistent with the Figure 3 phase diagram.  Note that a fixed-B hysteresis curve measured at B = 5 T and displayed in the inset also indicates no obvious hysteresis.  For {\bf B} $\parallel b$, however, measurements carried out at T = 46 and 47 K behave similarly to each other, and quite differently than the {\bf B} $\parallel a$ measurement.  That is, the transitions occur over a broad field range and they are irreversible, as $\Delta c$ is 0.01--0.02\% as the magnetic field is ramped down toward 0 T.  This irreversibility may be related to the first order nature of the field-induced transition for this orientation.  It should be noted that a flopside or spin-flop magnetic structure with moments aligned transverse to the applied magnetic field has been proposed based upon magnetization\cite{mccall} and Raman scattering studies,\cite{karpus} and a transition between the antiferromagnetic phase and the spin-flop phase is expected to be first order and possibly hysteretic.\cite{dejongh}  Structural irreversibility seen in our data would then follow from strong magnetoelastic coupling in this material.\cite{singh}  A spin-flop magnetic phase might also explain the path dependence in phase space that was mentioned earlier.  That is, in the fixed-B measurement the T = 44.3 K and B = 8 T point in phase space is reached from within the spin-polarized phase, while in the fixed-T measurement the path includes the proposed spin-flop phase.  The spin-flop to spin-polarized transition is expected to be second order,\cite{dejongh} and therefore some of the magnetic moments may remain aligned along {\it a} at B = 8 T.

A comparison of our structural measurements with the reported transport and magnetization measurements now provides a more complete picture of the behavior of Ca$_3$Ru$_2$O$_7$ near T$_{m-i}$.  Just below $T_{m-i}$, magnetic field applied along {\it a} drives the spin reorientation, and via strong magnetoelastic coupling, the c-axis lattice parameter increases.  The ``colossal'' decrease in the resistivity follows from the charge-lattice coupling that ties together the expanded c-axis lattice parameter and metallic interlayer transport.  For magnetic field applied along {\it b}, two different phases compete in the vicinity of T$_{m-i}$.  One of these phases is the spin-polarized phase above the metamagnetic transition.  It does not couple to the lattice, and the moderate reduction in the resistivity exhibited by this phase can be attributed to tunneling aided by ferromagnetic interlayer coupling.\cite{cao3,yoshida1}  The other phase is the antiferromagnetic metallic phase with magnetic moments along {\it a} (i.e., above the spin reorientation transition), which does couple to the lattice, as described above.  This phase exhibits an expanded c-axis lattice parameter and low resistivity.  As T$_{m-i}$ is approached from below, these two phases compete and the result appears to be a compromise--- a phase with a slightly expanded c-axis lattice parameter and a moderate decrease in the resistivity.  Our measurements suggest that the c-axis lattice parameter in this phase is unaffected by increasing field, but that it grows gradually with increasing temperature.  This behavior is consistent with the proposed spin-flop magnetic structure.  X-ray or neutron scattering measurements would be worthwhile to pursue in order to verify this magnetic structure, as well as its evolution with temperature and magnetic field.

In conclusion, we have used high-field x-ray scattering techniques to characterize the lattice degree of freedom of Ca$_3$Ru$_2$O$_7$ near the metal-insulator transition.  The anisotropic behavior of the structural change for magnetic field applied along the {\it a} and {\it b} in-plane directions results in a phase diagram similar to that reported from transport measurements,\cite{mccall,lin} and indicates strong charge-lattice coupling in this material.  An intriguing phase arises for {\bf B} $\parallel$ {\it b}, in which competition between the spin-polarized phase above the field-induced metamagnetic transition, and the metallic antiferromagnetic phase that occurs above the spin reorientation transition, gives rise to rich behavior that underscores the delicate balance of the charge, lattice, and spin degrees of freedom in Ca$_3$Ru$_2$O$_7$.

We gratefully acknowledge S.C. LaMarra's assistance with the magnet and B. Keimer's critical reading of the paper.  Use of the NSLS was supported by the U.S. DOE, Office of Basic Energy Sciences, under Contract No. DE-AC02-98CH10886.

\pagebreak

\begin{figure}
\includegraphics[width=8.51cm]{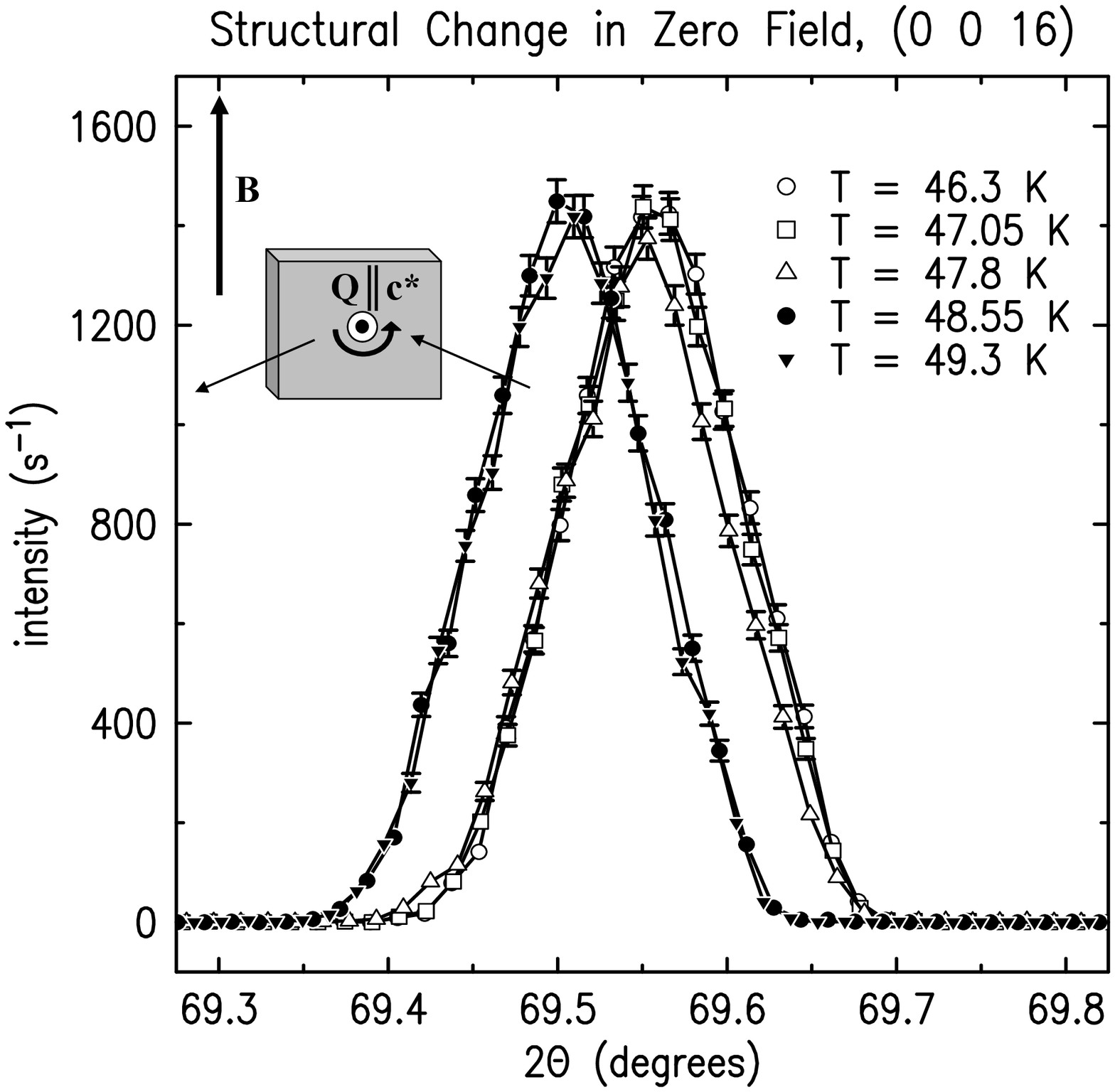}
\caption{Zero field $\theta$-$2\theta$ scans through the (0 0 16) reflection as a function of temperature. Inset displays the scattering geometry.}
\end{figure}

\begin{figure}
\includegraphics[width=6cm]{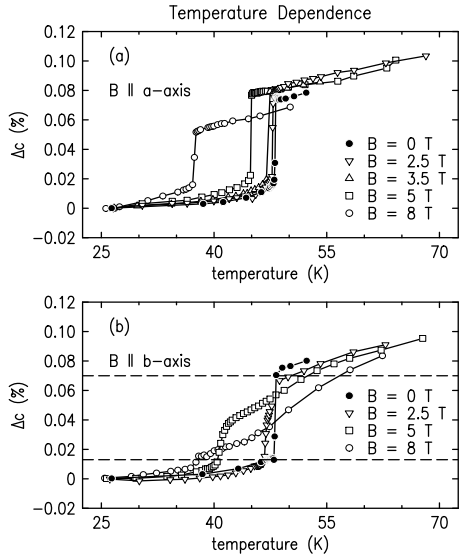}
\caption{Temperature dependence of the c-axis lattice parameter change--- relative to its value at T $\approx$ 25 K--- in fixed {\bf B} $\parallel$ {\it a}- (a) and {\it b}- (b) axis.  Dashed lines in (b) are the limits used to determine the temperature range of the structural change (see Figure 3).}
\end{figure}

\begin{figure}
\includegraphics[width=6cm]{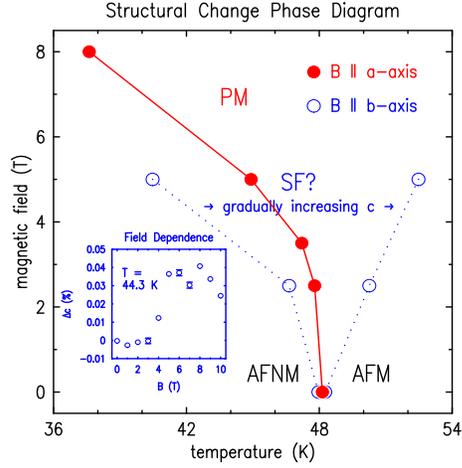}
\caption{Structural change phase diagram for {\bf B} applied along the {\it a}- ($\bullet$) and {\it b}- ($\circ$) axis, and the lines are merely guides for the eye.  Labels, which are taken from reference 7, are for the sake of comparison, and refer to paramagnetic metallic (PM), spin-flop (SF), antiferromagnetic nonmetallic (AFNM), and antiferromagnetic metallic (AFM) phases.  Inset displays the field dependence of the c-axis lattice parameter change at T = 44.3 K for {\bf B} $\parallel b$.}
\end{figure}

\begin{figure}
\includegraphics[width=8.62cm]{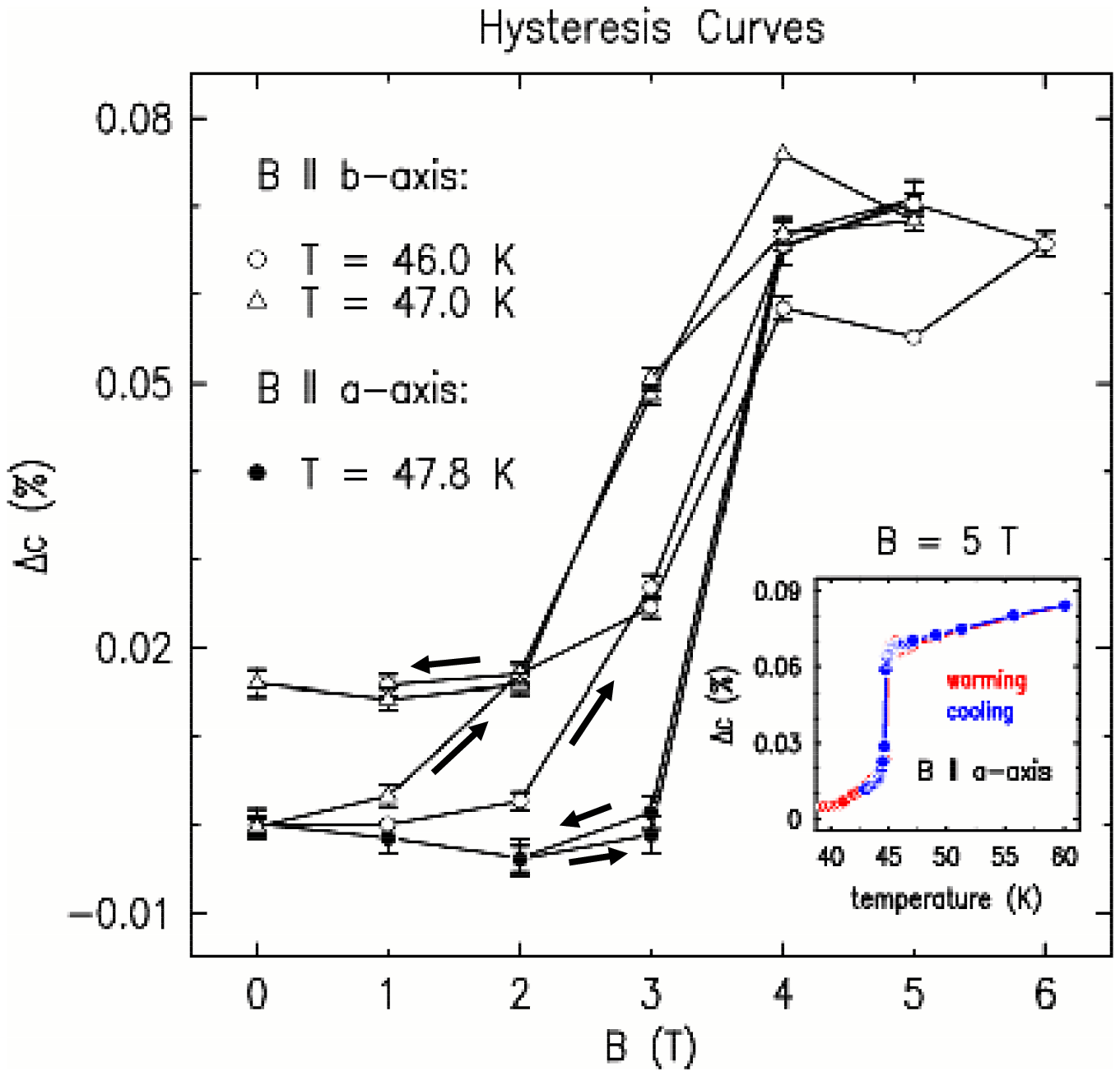}
\caption{Hysteresis curves as a function of {\bf B} applied along the {\it b}- ($\circ$, $\triangle$) and {\it a}- ($\bullet$) axis.  Inset displays the hysteresis curve for fixed {\bf B} (= 5 T) $\parallel a$.}
\end{figure}


\begin{thebibliography}{99}

\bibitem{kaplan} See, for example, {\it Physics of Manganites}, edited by T.A. Kaplan and S.D. Mahanti (Kluwer Academic, 1999).

\bibitem{bishop} A.R. Bishop, Synth. Met. {\bf 86}, 2203 (1997).

\bibitem{cao1} G. Cao {\it et al.}, Phys. Rev. Lett. {\bf 78}, 1751 (1997).

\bibitem{yoshida1} Y. Yoshida {\it et al.}, Phys. Rev. B {\bf 69}, 220411(R) (2004).

\bibitem{cao3} G. Cao {\it et al.}, Phys. Rev. B {\bf 67}, 060406(R) (2003).

\bibitem{yoshida2} Y. Yoshida {\it et al.}, Phys. Rev. B {\bf 72}, 054412 (2005).

\bibitem{mccall} S. McCall, G. Cao, and J.E. Crow, Phys. Rev. B {\bf 67}, 094427 (2003).

\bibitem{bohnenbuck} B. Bohnenbuck {\it et al.}, unpublished.

\bibitem{cao4} G. Cao {\it et al.}, Phys. Rev. B {\bf 67}, 184405 (2003).

\bibitem{abswitch} Cao and co-workers report {\it a} as the easy axis at low temperatures.  However our results, as well as resonant x-ray magnetic scattering studies of St. Andrews-grown samples, are consistent with the results of Yoshida {\it et al.}, Phys. Rev. B {\bf 72}, 054412 (2005), with {\it b} as the low-temperature easy axis.

\bibitem{lin} X.N. Lin {\it et al.}, Phys. Rev. Lett. {\bf 95}, 017203 (2005).

\bibitem{cao5} G. Cao {\it et al.}, Phys. Rev. B {\bf 69}, 014404 (2004).

\bibitem{cao2a} G. Cao {\it et al.}, Phys. Rev. B {\bf 62}, 998 (2000).

\bibitem{ohmichi} E. Ohmichi {\it et al.}, Phys. Rev. B {\bf 70}, 104414 (2004).
\bibitem{karpus} J.F. Karpus {\it et al.}, Phys. Rev. Lett. {\bf 93}, 167205 (2004).

\bibitem{dejongh} L.J. de Jongh and A.R. Miedema, Adv. Phys. {\bf 50}, 947 (2001).

\bibitem{singh} D.J. Singh and S. Auluck, Phys. Rev. Lett. {\bf 96}, 097203 (2006).
\end{thebibliography}
\end{document}